\documentclass[showpacs,preprintnumbers,amsmath,reprint,aps,prl,superscriptaddress,twocolumn]{revtex4-2}

\usepackage{bm}
\usepackage{graphicx}
\usepackage{color}
\usepackage{amsmath}
\usepackage{hyperref}
\usepackage{epstopdf}
\usepackage{upgreek}
\usepackage{mathptmx, textcomp}
\usepackage[latin1]{inputenc}
\usepackage[T1]{fontenc}
\usepackage{array, multirow}
\usepackage[table]{xcolor}
\usepackage{booktabs}
\usepackage{longtable}
\usepackage{comment}

\usepackage[normalem]{ulem}

\newcommand{\commentOut}[1]{}

\newcommand{\vib}{\ensuremath{\text{v}}}

\begin{document}

\title{
Controlling few-body reaction pathways using a Feshbach resonance
}

\author{Shinsuke Haze} \email{shinsuke.haze.qiqb@osaka-u.ac.jp}
\affiliation{Institut f\"{u}r Quantenmaterie and Center for Integrated Quantum Science and Technology IQ$^{ST}$, Universit\"{a}t Ulm, D-89069 Ulm, Germany}

\author{Jinglun Li}
\affiliation{Institut f\"{u}r Quantenmaterie and Center for Integrated Quantum Science and Technology IQ$^{ST}$, Universit\"{a}t Ulm, D-89069 Ulm, Germany}

\author{Dominik Dorer}
\affiliation{Institut f\"{u}r Quantenmaterie and Center for Integrated Quantum Science and Technology IQ$^{ST}$, Universit\"{a}t Ulm, D-89069 Ulm, Germany}

\author{Jos\'{e} P. D'Incao}
\affiliation{Institut f\"{u}r Quantenmaterie and Center for Integrated Quantum Science and Technology IQ$^{ST}$, Universit\"{a}t Ulm, D-89069 Ulm, Germany}
\affiliation{JILA, NIST and Department of Physics, University of Colorado, Boulder, CO 80309-0440, USA}

\author{Paul S. Julienne}
\affiliation{Institut f\"{u}r Quantenmaterie and Center for Integrated Quantum Science and Technology IQ$^{ST}$, Universit\"{a}t Ulm, D-89069 Ulm, Germany}
\affiliation{Joint Quantum Institute, University of Maryland and NIST, College Park, MD 20742, USA}

\author{Eberhard Tiemann}
\affiliation{Institut f\"ur Quantenoptik, Leibniz Universit\"at Hannover, 30167 Hannover, Germany}

\author{Markus Dei{\ss}}
\affiliation{Institut f\"{u}r Quantenmaterie and Center for Integrated Quantum Science and Technology IQ$^{ST}$, Universit\"{a}t Ulm, D-89069 Ulm, Germany}

\author{Johannes Hecker Denschlag} \email{johannes.denschlag@uni-ulm.de}
\affiliation{Institut f\"{u}r Quantenmaterie and Center for Integrated Quantum Science and Technology IQ$^{ST}$, Universit\"{a}t Ulm, D-89069 Ulm, Germany}

\date{\today}

\begin{abstract}
Gaining control over chemical reactions on the quantum level is a central goal of the modern field of cold and ultracold chemistry.  
Here, we demonstrate a novel method to coherently 
steer reaction flux of a three-body recombination process across different product spin channels. For this, we employ a magnetically-tunable Feshbach resonance
to admix, in a controlled way, a specific 
spin state to the reacting collision complex.  
This allows for the control of the reaction flux
into the admixed spin channel, which can be used to significantly change the reaction products.
Furthermore, we also investigate the influence of an Efimov resonance on the reaction dynamics.
We find that while the Efimov resonance can be used to globally enhance three-body recombination, the relative flux between the reaction channels
remains unchanged. 
Our control scheme is general and
can be extended to other reaction processes.
It also provides new opportunities in combination with other control schemes, such as quantum interference of reaction paths.
 
\end{abstract}

\maketitle

A chemical reaction in a low-density 
gas phase is typically well-described by 
a fully coherent quantum mechanical evolution. 
Therefore,  such a 
gas  is a promising testbed for quantum control of chemical processes. 
In fact, recent platforms based on ensembles of ultracold atoms or molecules have paved the way for extended quantum mechanical steering of reactions. Demonstrated control schemes include the use of photoassociation \cite{Jones2006,Lett1993,Miller1993}, Feshbach resonances \cite{Donley2002,Herbig2003,Weber2003,Yang2022,Chin2010,Park2023b}, microwave-engineered collisions \cite{Anderegg2021,Lin2023, Bigagli2023,Chen2024,Yan2020}, electric-field-controlled reactions \cite{Matsuda2020}, relative positioning of traps \cite{Ruttley2023,Yu2021,Reynolds2020,Cheuk2020}, confinement-induced effects \cite{Tobias2022,deMiranda2011,Drews2017,Goban2018,Sala2013,Lee2023}, quantum interference \cite{Son2022,Liu2023}, or rely on propensity rules and conservation laws \cite{Hu2021,Haze2022}. This progress has been further promoted by emerging technologies that enable state-to-state measurements
(e.g. \cite{Wolf2017,Wolf2019,Liu2021,Rui2017,Hoffmann2018}). 

A prominent tool for controlling chemical reactions is a tunable Feshbach resonance.
A Feshbach resonance in atomic gases occurs when the energy of the scattering state of two colliding
atoms is tuned into degeneracy with that of a molecular state, leading to the mixing of two such states \cite{Chin2010}. As they offer unique control over the interparticle interaction, tunable Feshbach resonances have been essential for the development of the field of ultracold quantum gases. 
An established application of Feshbach resonances for chemical reactions is the controlled production of ultracold molecules. By magnetically ramping over a Feshbach resonance, ultracold pairs of atoms can be converted into an extremely weakly-bound molecule, the Feshbach molecule \cite{Donley2002,Herbig2003,Duerr2004,Xu2003,Thalhammer2006,Chin2010}. In three-body recombination where three free atoms collide to form a diatomic molecule, Feshbach resonances have been used to tune the total molecular production rate
and specifically to suppress atom loss
\cite{Jochim2003,Xie2020, Braaten2006, Weber2022}, and to demonstrate the Efimov effect \cite{Ferlaino2011,Braaten2006,dIncao2018}. 
Feshbach resonances and resonant scattering
have also been proposed for controlling
complex few-body reactions, see e.g.
\cite{Hermsmeier2021,Tscherbul2015}. 

Here, we demonstrate the use of a Feshbach resonance 
in a three-body recombination process 
to steer reaction flux between families of molecular product channels with different spin states.  More specifically, by tuning the
magnetic field towards a Feshbach resonance we can gradually increase the initially negligible reaction rate into a specific spin channel, 
so that it becomes close to the total rate into all channels. 
The process is coherent and represents a novel tool for 
state-selective controlling of molecular production rates, using the applied magnetic field as a precisely tunable control knob. 

The experiments are carried out with 
a $860\:\textrm{nK}$-cold cloud of about $2.5 \times 10^5$ $^{85}$Rb atoms 
where each atom $i$ is 
in the hyperfine state 
$(f_i,m_{f_i}) = (2,-2)$ 
of the electronic ground state. The atoms are confined in a far-detuned crossed optical dipole trap, for more details see Methods and \cite{Haze2023}.
In the atom cloud, three-body recombination spontaneously occurs,
predominantly producing weakly-bound molecules in states of the coupled molecular complex $X^1\Sigma_g^+-a^3\Sigma_u^+$.
By tuning the magnetic field $B$ in the vicinity
of the  $s$-wave  Feshbach resonance
at $B = 155\:\textrm{G}$ \cite{Blackley2013,Koehler2006}
we can control the product distribution of the molecules. For the details of the Feshbach resonance, see also Supplemental Materials.
The molecules are state-selectively detected via resonance-enhanced multiphoton ionization (REMPI), see Methods for details.  

\begin{figure}[t]
	\includegraphics[width=0.7\columnwidth]{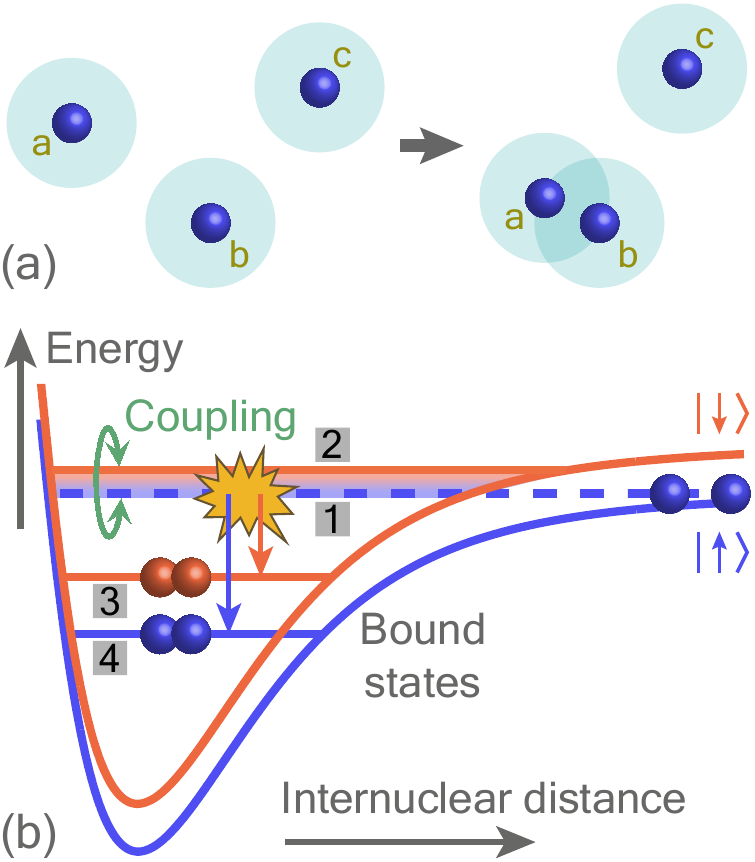}
	\caption{ Scheme for controlling the reaction flux into different spin channels using a two-body Feshbach resonance. (a) 
		Atoms ($a,b,c$) undergo three-body recombination, where ($a,b$) form a molecule.
		During this process, atom $(c)$ is outside the
		ranges (cyan areas) for spin-exchange interaction with the other atoms. 
		(b) Schematic representation of the Born-Oppenheimer potential energy curves for the atom pair ($a,b$). At close distance, the incoming scattering state (1)  with spin
		$|\hspace{-2pt}\uparrow \rangle$
		experiences admixing of the bound state (2)  which has spin 
		$|\hspace{-2pt}\downarrow \rangle$.
		Upon collision with the third atom ($c$) (not shown here) the 
		scattering state can then relax into  molecular bound states (3) or (4), with their respective spin states 
		$|\hspace{-2pt}\downarrow \rangle$ and $|\hspace{-2pt}\uparrow \rangle$.
	}
	\label{fig1new}
\end{figure}

Our scheme for controlling the reaction flux into different spin channels is  illustrated in Fig. \ref{fig1new}. In the three-body recombination process, the Rb atoms ($a,b,c$) collide and ($a,b$) 
form a molecule, see Fig. \ref{fig1new}(a). 
In the particular reactions we study here, the third atom ($c$) is far enough away, so that it interacts with the
atoms ($a,b$) merely mechanically and 
no spin flip between ($c$) and the pair
($a,b$) occurs \cite{Haze2022,note1}. 
Therefore, spin physics aspects of 
the reaction  can be understood to a large extent in a two-body picture, where  atom $(a)$ collides with atom $(b)$.
At large internuclear distances the ($a,b$) scattering state has the hyperfine spin quantum numbers 
$(F, f_a,f_b,m_F) = (4,2,2,-4)$,
where $F$ denotes the total angular momentum of the molecule excluding rotation, and  $m_F= m_{f_a} + m_{f_b}$ represents its projection.
We denote this spin state by $|\hspace{-3pt}\uparrow \rangle$.
At short internuclear distances the scattering state couples to an energetically near-by molecular bound level,
giving rise to the Feshbach resonance. 
This level has the spin state 
$(F,f_a,f_b,m_F) = (4,3,3,-4)$
which we denote by $|\hspace{-2pt}\downarrow \rangle$ \cite{Levelinfo}.
The coupling leads to an admixture of the $|\hspace{-2pt}\downarrow \rangle$ state to the initial scattering state with spin $|\hspace{-2pt}\uparrow \rangle$, and the strength of this admixture can be magnetically controlled. 
Next, in the mechanical collision with atom $(c)$, the scattering state of
($a,b$) can transition into a 
molecular bound state. 
Due to angular momentum conservation, 
the spin state of the newly formed molecule  must, however, have overlap either with the spin state  $|\hspace{-2pt}\uparrow \rangle$ or with 
$|\hspace{-2pt}\downarrow \rangle$
%\footnote{This argument is closely linked 
	%to a spin conservation propensity rule that was observed in recent work of ours \cite{Haze2022},
	\cite{conservation}.
	In fact, by tuning the $|\hspace{-2pt}\downarrow \rangle$
	admixture of the ($a,b$) scattering state we can control the reaction flux into molecular product channels with 
	spin $|\hspace{-2pt}\downarrow \rangle$.  
	
	\begin{figure}
		\includegraphics[width=0.9\columnwidth]{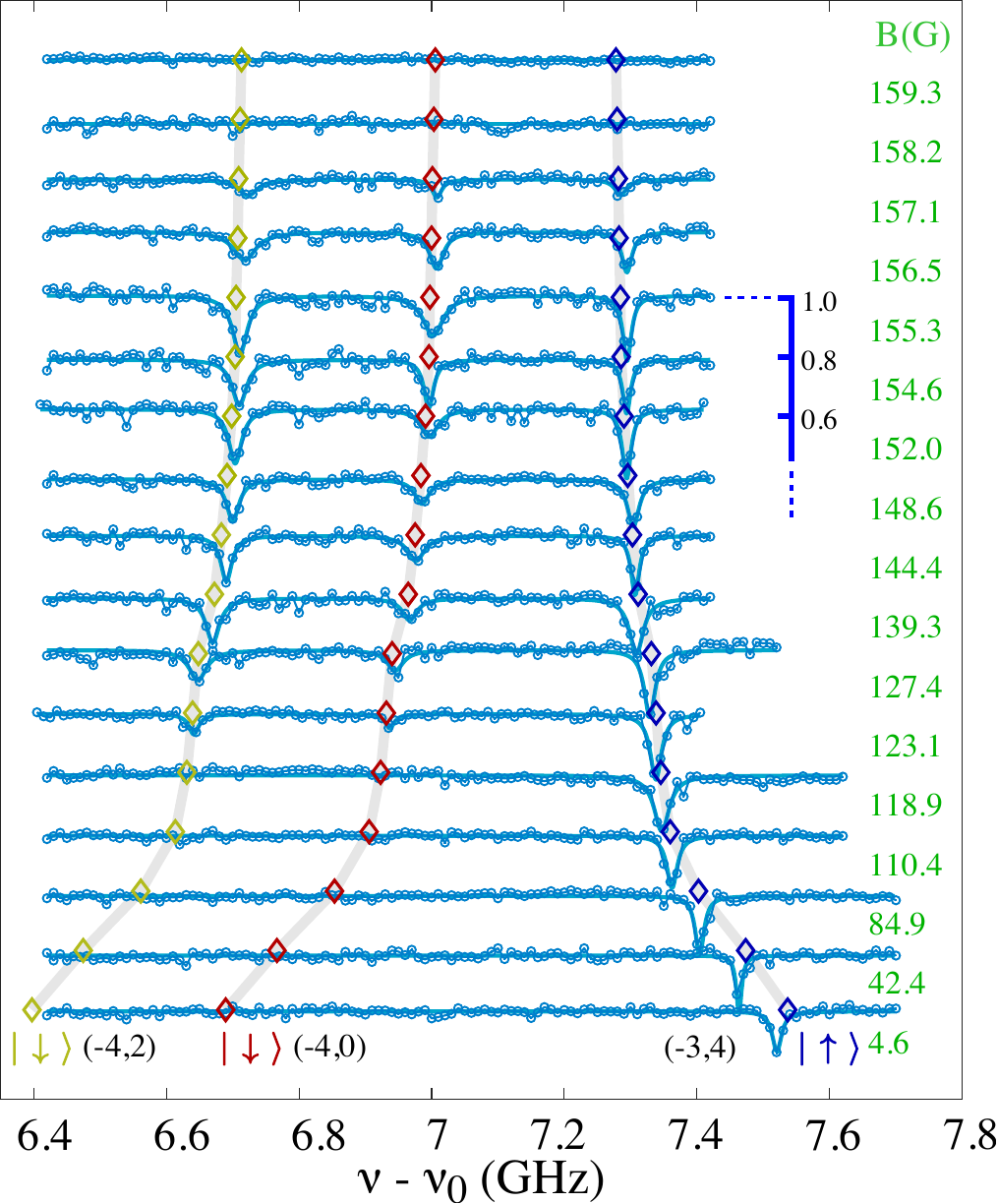}
		\caption{
Observation of $|\hspace{-2pt}\uparrow \rangle$ and $|\hspace{-2pt}\downarrow \rangle$ molecules. Shown are REMPI spectra as a function of the REMPI laser frequency $\nu$ for various magnetic fields $B$. Here, $\nu_0=497603.591\:\textrm{GHz}$. Each dip in a trace corresponds to a signal from a  distinct molecular level. The REMPI signals are 
  normalized, ranging from 0 to 1, as indicated by the 
  vertical bar. The bar is valid for all data traces. The diamonds mark the theoretical positions of possible molecular signals and the colors indicate the spin state as well as the vibrational and rotational level
			($\textrm{v}, L_R$). The faint color bands connecting the diamonds are guides to the eye.
			We note that the binding energy of the $|\hspace{-2pt}\uparrow \rangle$ level is smaller than that of the two 
			$|\hspace{-2pt}\downarrow \rangle$
			levels. 
			In the shown spectra, however, the
			signal for $|\hspace{-2pt}\uparrow \rangle$ is at higher frequency $\nu$ 
			since the intermediate rotational level for the REMPI is different,  see also Methods.
		}
		\label{fig:s1}
	\end{figure}
	
	We now demonstrate this control scheme experimentally. 
	Figure \ref{fig:s1} presents  REMPI spectra in a selected frequency range,
	showing signals from three different molecular product states. 
	The spectra  have been taken within a range of 
	magnetic fields $B$ between $4.6\:\textrm{G}$ and $159.3\:\textrm{G}$. (Note the nonuniform step size of $B$.)
	$\nu$ represents the REMPI laser frequency and 
	$\nu_0$ is a frequency reference,
	% given by a photoassociation resonance of two atoms at $B =$ 4.6 G, 
	see Methods.
	Each dip in a trace for a chosen $B$-field corresponds to a state-specific product molecule signal. Colored diamonds mark the known resonance positions predicted from coupled-channel calculations.
	
	\begin{figure*}
		\includegraphics[width=1.32\columnwidth]{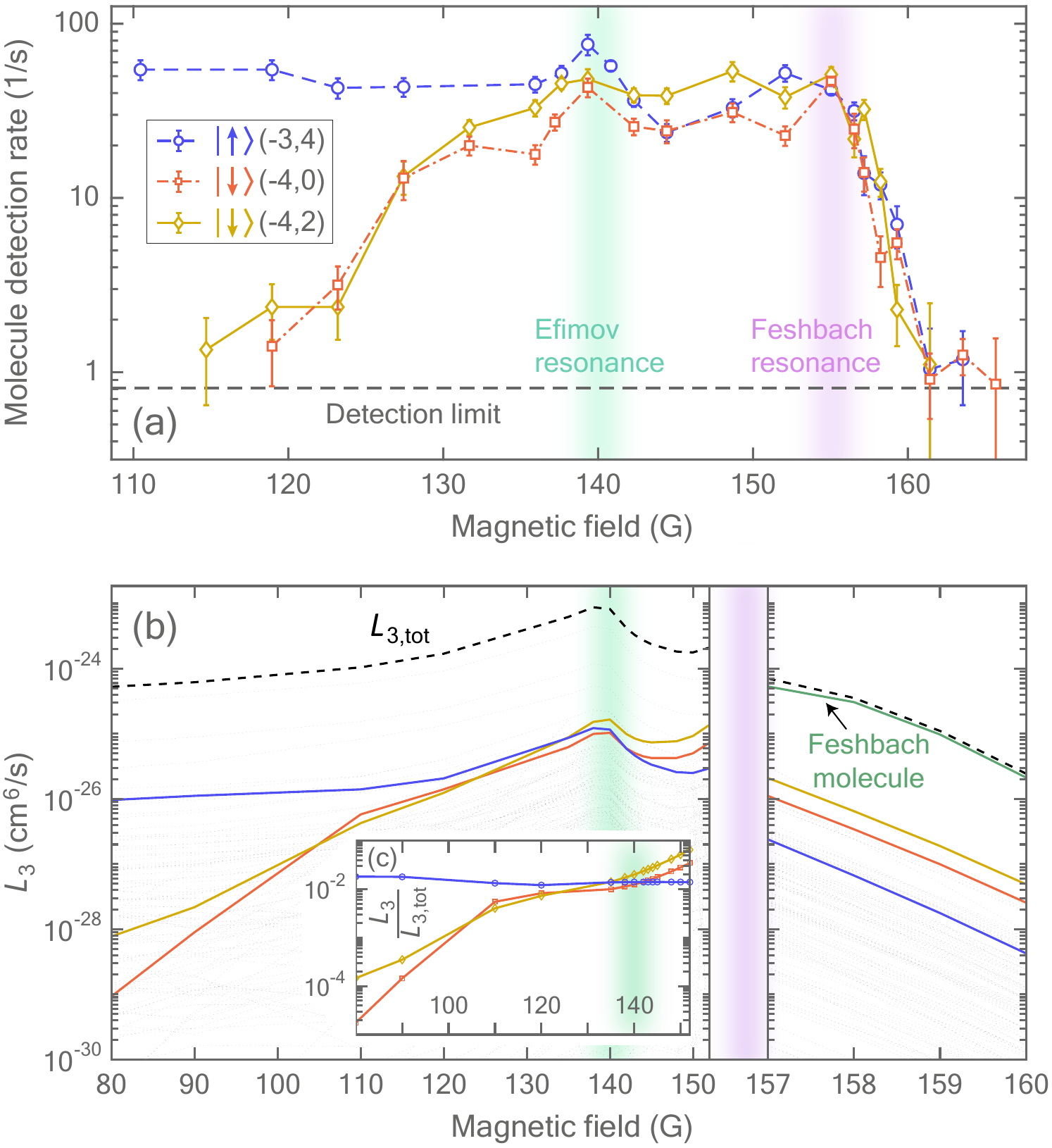}
		\caption{
			Opening up a product spin channel. (a) Molecule detection rates for three product states for which the quantum numbers $ | \uparrow / \downarrow \rangle  (\textrm{v}, L_R)$ are given in the legend.
			The gray dashed line marks the experimental detection limit and the light-green and purple shaded areas indicate the position of the Efimov and Feshbach resonance, respectively. The error bars in the plot indicate one standard deviation (1$\sigma$). (b) Calculated  three-body recombination rate coefficients.
			The black dashed line is the total rate coefficient $ L_{3,\mathrm{tot}} $.
			Solid colored lines correspond to partial rates for the states under discussion.  Gray lines correspond to other molecular states. 
			No calculations are shown for $152\:\textrm{G}\leq B \leq 157\:\textrm{G}$, see text.
   We expect theoretical 
   errors up to a few tens of percent for the partial rate 
   for vibrational levels down to v $= -4$, judging from when more vibrational states are included in our effective potentials \cite{Haze2023}.
			(c) The normalized reaction rate coefficients $L_3/L_{3,\mathrm{tot}}$ do not exhibit a maximum at the Efimov resonance. 
		}
		\label{fig2}
	\end{figure*}

	The molecular levels are labeled by
	their spin states  ($|\hspace{-2pt}\uparrow \rangle$ or $|\hspace{-2pt}\downarrow \rangle$), and 
	by their vibrational ($\textrm{v}$) and  rotational ($L_R$) quantum numbers
	\cite{Koehler2006}. 
	In Fig. \ref{fig:s1} we observe one molecular state with spin  $|\hspace{-2pt}\uparrow \rangle$ 
	and two 
	molecular states with spin $|\hspace{-2pt}\downarrow \rangle$. 
	The resonance positions %of the molecular states 
	change in a characteristic way with the $B$-field due to the Zeeman effect.
	We use this as 
	fingerprint information for identifying the molecular levels.
	The strength of each signal roughly reflects the  recombination rate towards each respective state. 
	In the magnetic field range up to about $120\:\textrm{G}$ each REMPI spectrum exhibits only a single resonance dip which can be unambiguously assigned to the state ${|\hspace{-2pt}\uparrow \rangle}$. 
	At about $120\:\textrm{G}$, two additional molecular signals start to appear 
	stemming from molecular states in the spin state
	$|\hspace{-2pt}\downarrow \rangle$. 
	The strengths of the signals of these states become similar to the $|\hspace{-2pt}\uparrow \rangle$ signals when approaching the Feshbach resonance at $155\:\textrm{G}$. For magnetic fields above the Feshbach resonance, all signals decrease very quickly within a few Gauss.
	
	From our REMPI spectra, 
	molecule detection rates for each observed molecular state are extracted. 
	These rates are roughly proportional to the 
	partial three-body recombination rates for the flux into individual product channels (see \cite{Haze2023} and Methods).  The  obtained rates for the  states in Fig.~\ref{fig:s1} are shown in Fig. \ref{fig2}(a) for the magnetic field region in the vicinity of the Efimov and Feshbach resonances, located at $140\:\textrm{G}$ and $155\:\textrm{G}$, respectively. 
	
	The data show that the rates
	for the $|\hspace{-2pt}\downarrow \rangle$ states indeed strongly increase from below the detection limit (gray dashed line) to about a factor of 50 above the detection limit as the magnetic field is increased from  
	$B<115\:\textrm{G}$  towards
	the Efimov resonance.
	The detection limit is mainly determined by the background noise of our REMPI scheme. 
	By contrast, the rate for the
	$|\hspace{-2pt}\uparrow \rangle$ state is rather constant for all $B$-fields
	below the Efimov resonance. 
	At the position of the Efimov resonance at 140 G we observe  a clear enhancement of the rates for all molecular states. In fact, the signals for all three states attain similar strength, which demonstrates the large relative tuning range of our scheme. At this point the spin product channel $|\hspace{-2pt}\downarrow \rangle$ has been fully opened up for the reaction flux.
	
	The relatively constant production rate for the 
	$|\hspace{-2pt}\uparrow \rangle$ state below the Efimov resonance
	might be unexpected at first in view of the known $a^4$ scaling of the recombination rate in the limit of zero temperature \cite{Fedichev1996,Weber2003}, where $a$ is the scattering length.  
	It can, however, be explained to a large extent as an effect of our finite temperature of 860 nK
	\cite{Incao2004,Rem2013}, as further discussed below.

	We carried out numerical model calculations for the
	partial rate coefficients $L_3$ for each 
	molecular quantum state,  using the adiabatic hyperspherical representation \cite{wang2011pra,Chapurin2019,Xie2020} (see Methods). 
	This determines the 
	partial recombination rate, $ L_3(f) \cdot \int n^3 d^3r \, /3$, into a
	molecular state $f$, where $n$
	is the atomic density distribution. 
	The
	calculations take into account thermal averaging of the partial rate constants.
	The  results are shown in Fig.~\ref{fig2}(b).
	The region from $152$ to $157\:\textrm{G}$, i.e., the direct vicinity of the Feshbach resonance is excluded since  the numerical  calculations quickly become computationally highly demanding  in this resonant regime \cite{comment2}.

	Among all the possible molecular states produced by recombination (gray dotted lines), we highlight in color the molecular states 
	under discussion.  
	In addition, we present the total three-body recombination rate coefficient (dashed black lines). 
	
	Our calculations show that due to thermal averaging, the calculated recombination rate coefficient for the probed $|\hspace{-2pt}\uparrow \rangle$ state increases only moderately towards the Efimov resonance. For more details on how finite temperature affects the recombination rate coefficient, see Supplemental Materials.
	The increase of the theoretical curves is still faster than for
	the experimental data. This may be mainly explained by 
 imperfections of the experiment. 
 During the
 $B$-field ramp atoms are already lost due to three-body recombination and
 the sample slightly heats up. As a consequence, the density of the atom cloud sinks. In addition, the B-field ramp is not perfect, but tends to lag behind and to overshoot, which can lead to averaging out of signals. Furthermore, there could be a small variation in the REMPI efficiency as a function of magnetic field. 
 %\textcolor{blue}{\cite{comment3}}
These variations hamper a direct comparison between
 ion rates and $L_3$ coefficients.
	
Nevertheless,
 the main characteristics of the experimental data are qualitatively well described. For example, the observed sharp drop of the recombination rate above the Feshbach resonance is also clearly reproduced by the theory.
	The reason for this drop is the rapid decrease of the scattering length towards its
	zero crossing near 
	$B=166\:\textrm{G}$ and the close-by
	minimum in $L_3$ due to Efimov physics \cite{Xie2020,Zaccanti2009}.
    We note that above the Feshbach resonance the Feshbach molecular state appears (see dark-green solid line in Fig. \ref{fig2}(b)), and takes the main fraction of the total reaction flux.
	
	Our calculations show that the effect of the Efimov resonance is to increase the partial three-body recombination rate coefficients with the same overall factor, not favoring particular product channels. 
	This is evident from Fig. \ref{fig2}(b)
	where all partial rate coefficients exhibit a similar maximum at the location of the Efimov resonance. It also becomes manifest when normalizing the partial rate coefficients to the total rate coefficient (see Fig. \ref{fig2}(c)), 
	as each maximum 
	at the Efimov resonance disappears.
	The global enhancement is due to the fact  that the Efimov resonance is a shape resonance which occurs in a single three-body adiabatic channel. 
	As such,  approaching the resonance increases the overall amplitude of the three-body scattering wavefunction at short distances where the reaction takes place, therefore, enhancing all the partial rates by the same factor \cite{dIncao2018}.
	
	\begin{figure}
		\includegraphics[width= 0.9\columnwidth]{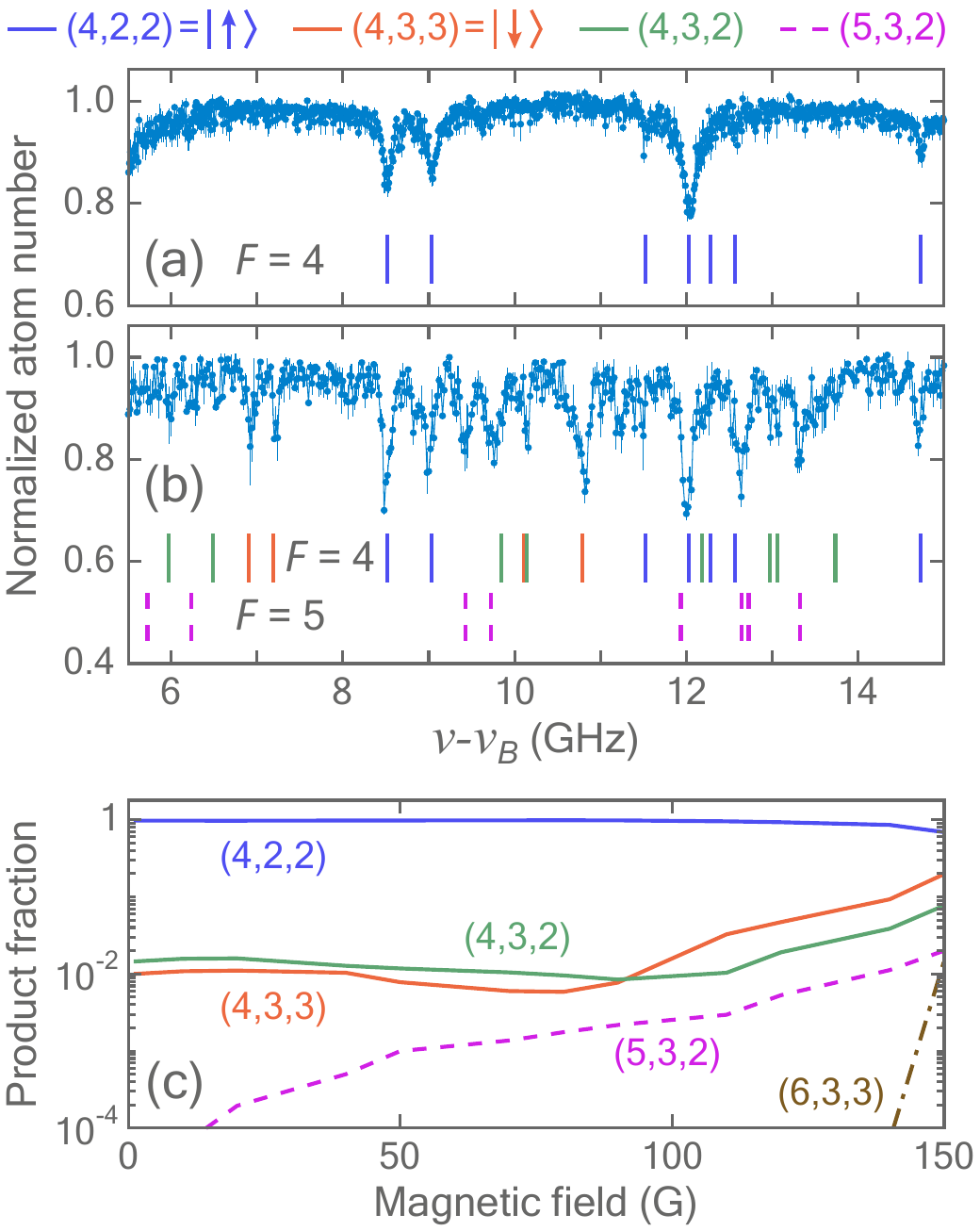}
		\caption{Spin families of molecular products. (a) and (b) Detection signals of product molecules of various spin families by REMPI spectroscopy at a magnetic field of $B=4.6\:\textrm{G}$ and $B=155\:\textrm{G}$, respectively. Vertical lines correspond to calculated resonance positions for molecular states assigned to spin families ($F,f_a,f_b$) according to the legend on top of the figure. The marked individual states have vibrational quantum numbers in the range from $\textrm{v}=-1$ to $-7$ and rotational quantum numbers $L_R=0,2,4$ or $6$.  The REMPI path is via a $^3 \Pi_g$
			intermediate state, see Methods. We have $\nu_B = \nu_0 =
			497831.928$ GHz for (a) and $\nu_B =
			\nu_0 - 228$ MHz
			for (b). The 228 MHz shift compensates for the Zeeman shift, allowing for a better comparison of the two spectra.
			(c) Calculated, summed-up molecular product fraction for each spin family $(F,f_a,f_b)$ as indicated next to the curves. }
		\label{fig4}
	\end{figure}
	
	Remarkably, near the Feshbach resonance we also find, both experimentally and theoretically, molecular products in spin states other than $|\hspace{-2pt}\uparrow \rangle$ and 
	$|\hspace{-2pt}\downarrow \rangle$,  as shown in Fig.~\ref{fig4}. This points towards physics beyond  the $|\hspace{-2pt}\uparrow \rangle$-$|\hspace{-2pt}\downarrow \rangle$ Feshbach mixing. 
	In the experiments, these are molecular products with spins $(F,f_a,f_b)=(4,3,2)$ 
	and $(5,3,2)$. In this notation, we omit $m_F$, as it is always $m_F = -4$. 
	Thus, we observe product 
	states where only one of the $f_i$ has
	flipped, and where even the total angular momentum $F$ can change. 
	Following a similar analysis as used in Refs.~\cite{Haze2022,Haze2023}, the observation of such states can be understood as follows in terms of two-body physics. 
	The spin state $(4,3,2)$ can be produced via two-body spin-exchange interaction at short distances, starting either from state $|\hspace{-2pt}\uparrow \rangle\equiv (4,2,2)$ or  from
	$|\hspace{-2pt}\downarrow \rangle\equiv(4,3,3)$. 
	Close to the Feshbach resonance the scattering wavefunction amplitude is strongly enhanced at short range and with it also the rate for spin-exchange.
	Producing the spin state $(5,3,2)$ is possible due to the presence of a finite $B$-field which breaks global rotational symmetry and couples different $F$ quantum numbers. A $(5,3,2)$ state
	is typically energetically close to a corresponding $(4,3,2)$ state, such that
	coupling between them is resonantly enhanced. 
	
	In Fig.~$\:$\ref{fig4}(a) and (b) REMPI spectra at magnetic fields of $B=4.6\:\textrm{G}$ and $B=155\:\textrm{G}$, respectively, are compared.
	The spectrum at high magnetic field exhibits many more resonance lines than the spectrum at low field. 
	In a thorough analysis of the spectra, similarly as in Ref. \cite{Wolf2017}, we identified a total of four spin families for $B=155\:\textrm{G}$ and only a single one for $B=4.6\:\textrm{G}$. The individual spin states are marked with colored bars in Fig.$\:$\ref{fig4}(a) and (b). 
	
	Figure \ref{fig4}(c) shows our numerical calculations for the product fractions of the molecules in the different spin families $(F,f_a,f_b)$ as a function of the external magnetic field $B$. Here, for each individual family we sum up the populations for all corresponding molecular states having the same spin characteristics. In agreement with our previous discussion,  spin-exchange is strongly enhanced 
	when approaching the Feshbach resonance. Furthermore, 
	Fig. \ref{fig4}(c) reveals a hierarchy in the propensity for the production of spin states. Changing the total 
	angular momentum $F$ is more strongly suppressed than changing an atomic $f$ quantum number, see also \cite{Lipaper}.
	
	In summary, we have demonstrated a powerful scheme to control the reaction pathway in a three-body recombination process of ultracold atoms. Using a magnetically tunable Feshbach resonance we 
	admixed  a well-defined  spin  state to the reaction complex of three atoms and by this steered
	%the size of the admixture determined
	the reaction flux between the corresponding spin channels. We find that a large fraction of the total reaction flux can be redirected in this way. 
	Furthermore, we show that in contrast to the Feshbach resonance an Efimov resonance only enhances globally the reaction rate,
	while maintaining the relative flux between reaction channels. 
	We investigated our control scheme
	both experimentally and theoretically, using
	high-resolution state-to-state measurements and 
	state-of-the-art numerical three-body scattering calculations, respectively.

	The demonstrated reaction control holds large promise for general few-body reactions, as it is simple and can easily be extended. Feshbach resonances are ubiquitous in cold atomic and molecular gases. The scheme is fully coherent and can thus be used as a central building block in interferometric control, where the Feshbach resonance functions as a beam splitter for the incoming wave function. The split up parts can then potentially follow different pathways towards the same final product state
 where they interfere. For example, the final product state could have tunable spin-mixed character which can be set by
  further control methods such as state dressing with optical or microwave fields. In this way, additional tuning of the interference can be achieved.

\section{acknowledgements}  
This work was financed by the Baden-W\"{u}rttemberg Stiftung through the Internationale Spitzenforschung program (contract BWST ISF2017-061) and by the German Research Foundation (DFG, Deutsche Forschungsgemeinschaft) within contract 399903135. We acknowledge support from bwFor-Cluster JUSTUS 2 for high performance computing. J. P. D. also acknowledges partial support from the U.S. National Science Foundation, Grant No. PHY-2012125 and PHY-2308791, and NASA/JPL
1502690. S.H. also acknowledges support from Japan Science and Technology Agency Moonshot R$\&$D Grant No. JPMJMS2063 and ASPIRE Grant No. JPMJAP2319. J.H.D and J.P.D. also acknowledge funding by Q-DYNAMO (EU HORIZON-MSCA-2022-SE-01) within project No. 101131418.

\section{Author contributions}
S.H. and D.D. have carried out the experiments. J.L. and J.P.D. calculated the three-body recombination rate coefficients. J.H.D. supervised the project. All authors have contributed to the analysis of the experiment and to the writing of the manuscript.

\section{Competing interests}
The authors declare no competing interests.

\bibliography{refs}

\begin{thebibliography}{63}
\bibitem [{Lev()}]{Levelinfo}%
  \BibitemOpen
  \href@noop {} {\bibinfo {title} {Furthermore, the level
  %$|\hspace{-2pt}\downarrow %\rangle$ involved in the % {F}eshbach resonance
  has rotational angular momentum ${L_R} = 0$ and its vibrational quantum
  number is $\vib = -3$, counting down from the $f_a = f_b = 3$ atomic
  threshold, starting with $\vib = -1$ for the most weakly-bound state. {T}he
  vibrational levels for {$|\hspace{-2pt}\uparrow \rangle$} states are counted
  analogously, however, starting from the $f_a = f_b = 2$ atomic
  threshold.}}\BibitemShut {Stop}%
\bibitem [{com()}]{comment2}%
  \BibitemOpen
  \href@noop {} {\bibinfo {title} {For the scattering calculations we need to
  calculate the three-body potentials and diabatic couplings up to a
  hyperradius at least an order of magnitude larger than the scattering length,
  all at a high resolution in length. {T}herefore, performing scattering
  calculations become increasingly difficult as we approach the {F}eshbach
  resonance, where the scattering length diverges.}}\BibitemShut {Stop}%
\end{thebibliography}

%\newpage

\section{Methods}

\subsection{Preparation of ultracold atomic sample}

The experimental sequence starts with capturing $^{85}$Rb atoms in a magneto-optical trap. After a magnetic transport over 40cm
the atoms are subsequently loaded into an optical dipole trap where evaporative cooling is performed. They are then transported to the center of the Paul trap via a moving 1D-optical lattice. At the final stage of the sample preparation, the atoms are confined in a far-detuned crossed-dipole trap formed by 1064 $\:\textrm{nm}$ lasers. The trapping frequency is $\omega_{x,y,z} = 2\pi \times (156, 148, 18) \:\textrm{Hz}$. The resulting atom cloud consists of a pure sample in $(f_i,m_{f_i}) = (2,-2)$ hyperfine spin state with the typical particle number of 2.5 $\times$ 10$^5$. The temperature of atoms is 860 $\:\textrm{nK}$.
This temperature was chosen as it provided the strongest recombination signals at a reasonably cold temperature.

\subsection{REMPI detection}

In order to state-selectively detect the product molecules, we apply two-step resonance-enhanced multiphoton ionization (REMPI) with a cw-laser which has a linewidth of $\approx 1\:\textrm{MHz}$. The laser beam is roughly an equal mixture of $\sigma$- and $\pi$-polarized light. 
%can be essentially considered as unpolarized. 
It 
has a power of 100$\:$mW and a beam waist ($1/e^2$ radius) of 0.1mm at the location of the atom cloud. 
We use identical photons for the two REMPI steps at a wavelength around 602$\:$nm. 
For Figs. \ref{fig:s1} and \ref{fig2} the intermediate REMPI states are levels of $(2) ^1\Sigma^+_u$ with $J' = 3$ for $|\hspace{-2pt}\uparrow \rangle$ states and $J' = 1$ for $|\hspace{-2pt}\downarrow \rangle$ states \cite{Haze2022}, where $J'$ represents the total angular momentum excluding nuclear spin. The $J' = 1 $ and $J' = 3 $  levels are split by 2.9 GHz. 
The photoassociation laser frequency towards 
the intermediate 
level $J' = 1 $ is $\nu_0 = 497603.591$ GHz at
$B = 4.6\:\textrm{G}$.
The binding energies of the experimentally observed  molecular states in Figs. \ref{fig:s1} and \ref{fig2} 
are $4.7\:\textrm{GHz}\times h$
for 
$|\hspace{-2pt}\uparrow \rangle$,
and span a range between 6.4 to 7.3$\:\textrm{GHz}\times h$
for  $|\hspace{-2pt}\downarrow \rangle$.
Here, the binding energy is determined
relative to the $B$-field dependent (4,2,2) threshold. 
For Fig. \ref{fig4}, the intermediate REMPI states are deeply-bound levels of  $(2) ^3\Pi \, \, 0_g^+$ with $J' =1, 3, 5$ \cite{Haze2022}. Here, $\nu_0 = 497831.928$ GHz is the photoassociation frequency towards $J' =1$ at
$B = 4.6\:\textrm{G}$.
The binding energies of the  molecular states observed in Figs. \ref{fig4} (a) and (b)  span a range between $0.6$ and $12.6\:\textrm{GHz}\times h$. Again, the binding energy is determined
relative to the (4,2,2) threshold.  
In general, the Zeeman effects of  our intermediate states are negligible compared to the ones of the ground state.
We make an effort to ensure that the REMPI efficiencies are similar for the states that we probe, also at various magnetic fields, but a precise calibration of the
REMPI efficiency has not been done yet. We note that the first REMPI step is generally not saturated.

When ions are produced via REMPI, they are directly
trapped and detected in an eV-deep Paul trap which is centered on the atom cloud. Elastic atom-ion collisions inflict tell-tale atom loss while the ions remain trapped. From the atom loss which is measured via absorption imaging of the
atom cloud the ion number can be inferred, for details see \cite{Haze2022}.
From the ion numbers and the interaction time 
we obtain an ion production rate (i.e. the molecular detection rate) which is generally proportional to the
state-selective molecular production rate
and the three-body recombination loss rate constant. 

\subsection{Model calculations}

Our numerical simulations use the
adiabatic hyperspherical representation approach 
where the 
coordinates of three particles are given in terms of the hyperradius $R$ for the overall size of the system and a set of hyperangles $\Omega$ for the internal motion \cite{Suno2002,wang2011pra,Chapurin2019,Xie2020}. The three-body Schr\"odinger equation is solved by adiabatically separating the hyperradial motion

\begin{align}
&\left[-\frac{\hbar^2}{2\mu}\frac{d^2}{dR^2}+U_{\nu}(R)\right]F_{\nu}(R)\nonumber\\
&~~~~~~+\sum_{\nu'}W_{\nu\nu'}(R)F_{\nu'}(R)=EF_{\nu}(R),\label{Schro}
\end{align}
 from the internal motion
\begin{equation}
\hat{H}_{\rm ad}\Phi_{\nu}(R;\Omega)=U_{\nu}(R)\Phi_{\nu}(R;\Omega), \label{had}
\end{equation}

where the hyperradius $R$ appears only as a parameter. The diagonalization of the hyperangular adiabatic Hamiltonian $\hat{H}_{\rm ad}$ gives the three-body potentials $U_{\nu}$ and the channel functions $\Phi_{\nu}$, which are also used for computing the nonadiabatic couplings $W_{\nu\nu'}$, for the hyperradial equation. 

In our model, the hyperangular adiabatic Hamiltonian reads
\begin{eqnarray}
\hat{H}_{\rm ad}=\frac{\hat\Lambda^2(\Omega)+15/4}{2\mu R^2}\hbar^2+\sum_{\substack{i,j=a,b,c \\ i \neq j}}\hat{V}_{ab}(R,\Omega)+\sum_{i=a,b,c}\hat{H}^{\rm sp}_{i}(B),
\end{eqnarray}
where $\hat\Lambda$ denotes the hyperangular momentum operator \cite{Suno2002,wang2011pra} and $\mu = m/\sqrt{3}$ is the reduced mass of three identical atoms of mass $m$. The atomic spin Hamiltonian $\hat{H}^{\rm sp}_{i}$ for atom $i$ contains the  hyperfine  and Zeeman interaction, 
and to a very good approximation within the present work
its eigenstates are $|f_i,m_{f_i}\rangle$.
For two Rb atoms (e.g., $i$ and $j$) of the $5S_{1/2}+5S_{1/2}$ asymptote, the pairwise interaction $\hat{V}_{ij}$ can be expressed in terms of the  
electronic singlet  
and triplet 
Born-Oppenheimer potentials.
We use the potentials from Ref. \cite{Strauss2010} with an additional repulsive term $C/r^{12}_{ij}$ to reduce the number of bound states in our simulation.
Here, 
 $r_{ij}$ is the interatomic distance. 
Removing deeply bound states mitigates the computational hardship without affecting too much the results, as generally more deeply bound states play a less important role in the three-body recombination process \cite{Haze2023}. The truncation of the potentials 
 shall be explained in more detail in a separate publication \cite{Lipapertwo}. In brief, two parameters $C$ ($C_\mathrm{s}$ and $C_\mathrm{t}$) are adjusted individually for the truncated singlet and triplet potentials so that they contain
6 and 5 $s$-wave bound states, respectively, and so that
the known singlet and triplet scattering lengths are reproduced. Additional fine-tuning of the two $C$ parameters together with the atomic hyperfine splitting aims at reproducing the Feshbach resonance
at about 155 G.
As a result, the atomic hyperfine splitting is reduced by about $5 \%$ compared to the literature value. We use $C_\mathrm{s}= (0.3242030\:r_\mathrm{vdw})^6\cdot C_6$ and $C_\mathrm{t}= (0.3258900\:r_\mathrm{vdw})^6\cdot C_6$, where $r_\mathrm{vdw}= \frac{1}{2}(\frac{mC_6}{\hbar^2})^{1/4}$ is the van der Waals length and $C_6$ is the van der Waals coefficient. 

Interactions between the particles and with the external magnetic field $B$ couple various angular momenta. 
Therefore, the incoming spin channel 
$|2,-2\rangle|2,-2\rangle|2,-2\rangle$ can in principle be coupled to a range of spin channels
$|f_a,m_{f_a}\rangle|f_b,m_{f_b}\rangle|f_c,m_{f_c}\rangle$, where $f_i$ can be 2 or 3. 
We essentially only have the restriction that 
$M_{\rm tot}=m_{f_a}+m_{f_b}+m_{f_c}$ is conserved, as long as spin-spin interaction can be neglected. 
However, motivated by previous work \cite{Haze2022}, 
where we found a spin conservation propensity rule in three-body recombination of Rb atoms we restrict the  spin of the third atom to be  $|f_c,m_{f_c}\rangle=|2,-2\rangle$ in our calculations. One reason for this restriction could be that the third atom ($c$) interacts mainly mechanically with the other two, ($a,b$), while they are forming a molecule.
This approximation leads to a model of five coupled three-body channels with the quantum numbers
($F, f_{a}, f_{b}$) = (4, 2, 2), (4, 3, 2), (4, 3, 3), (5, 3, 2), and (6, 3, 3). 

\section{Supplemental Materials}

\subsection{Feshbach resonance at 155 Gauss}
The $s$-wave Feshbach resonance used in this work is located at 155.3 $\:\textrm{G}$. It  couples the incoming $(F, f_a,f_b,m_F) = (4,2,2,-4)$ state and the closed-channel bound state $(F, f_a,f_b,m_F) = (4,3,3,-4)$. 
The scattering length across the Feshbach resonance is well characterized by the relation $a(B)=a_{\textrm{bg}}(1-\frac{\Delta B}{B-B_{0}})$. Here, $a_{\textrm{bg}}=-443 a_{0}$, $\Delta B=10.9 \:\textrm{G}$, $B_{0}=155.3 \:\textrm{G}$ is the background scattering length, the width and the position of the resonance \cite{Blackley2013}. The scattering length is shown in Fig. S1.

\renewcommand{\figurename}{Fig. S}
\setcounter{figure}{0}

	\begin{figure}[h]
		\includegraphics[width= 0.9\columnwidth]{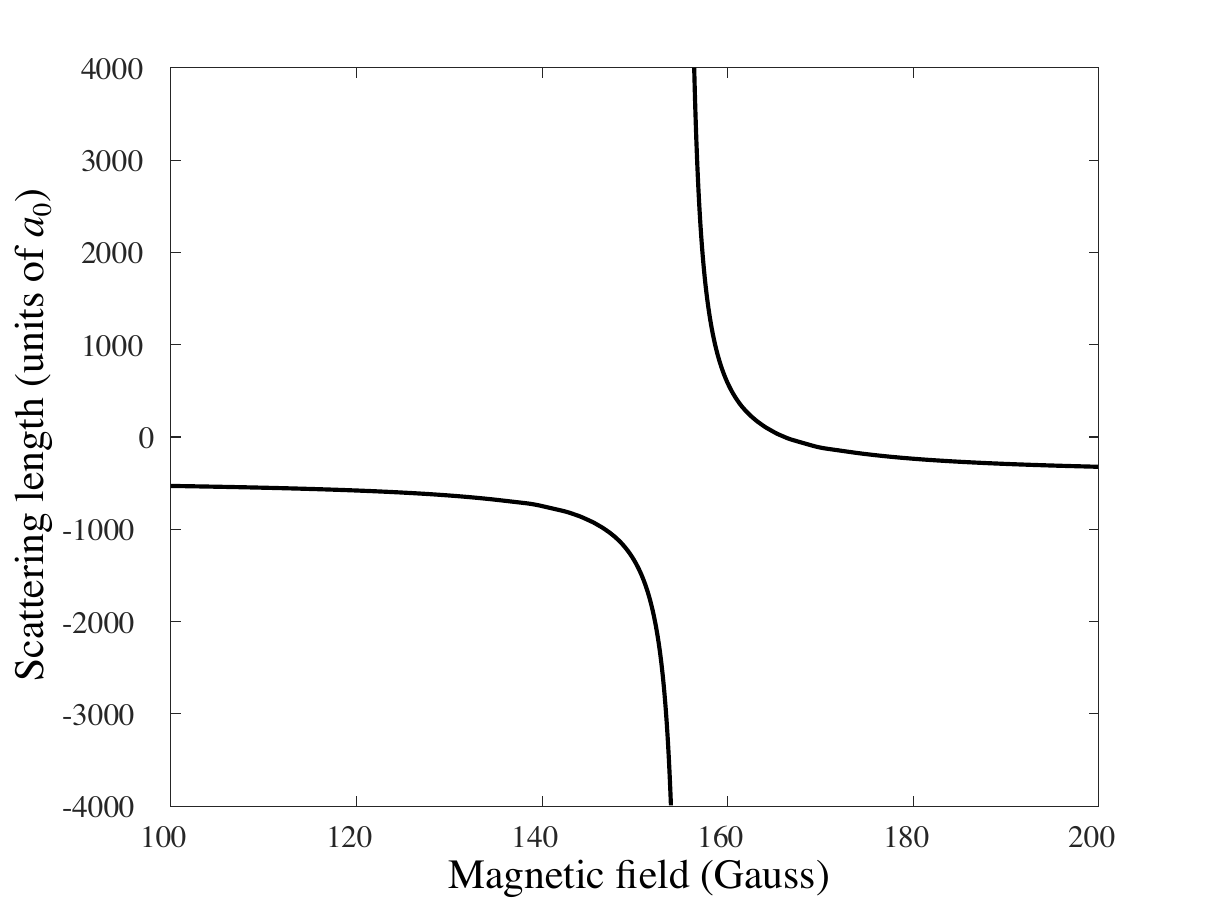}
		\caption{Scattering length in the vicinity of the $s$-wave Feshbach resonance. The scattering length in units of Bohr radius is plotted as a function of magnetic field.}
		\label{fig_SM1}
	\end{figure}
 
\subsection{Temperature dependence of $L_3$}
	\begin{figure}[h]
		\includegraphics[width= 1.0\columnwidth]{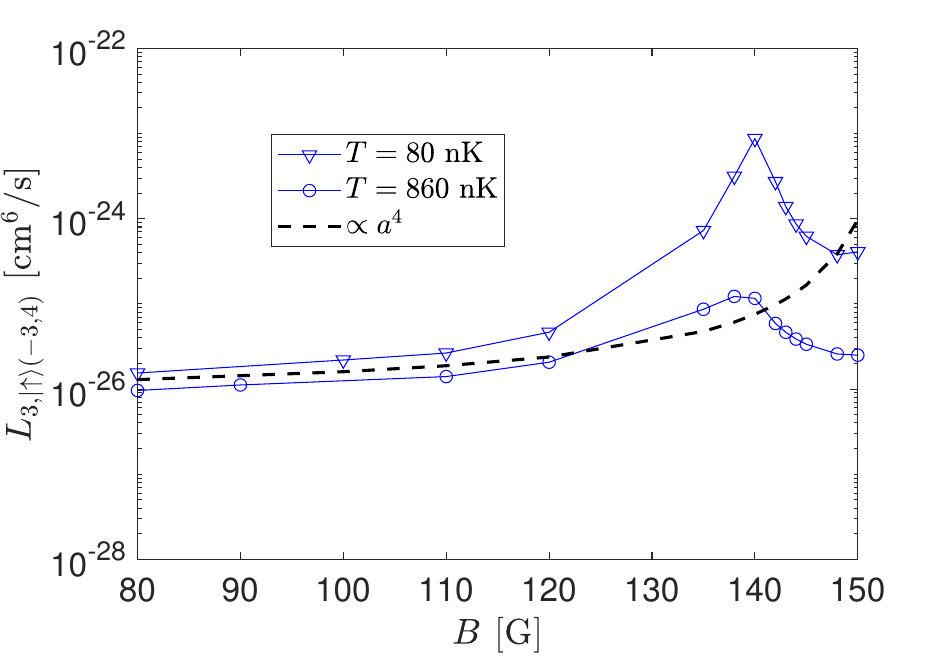}
		\caption{The three-body recombination rate constant $L_3$ for the $|\hspace{-2pt}\uparrow \rangle(-3,4)$ state at $80$ nK is compared to that at $860$ nK. The dashed line indicates the $L_3\propto a^4$ scaling.}
		\label{fig_SM2}
	\end{figure}
 The scaling of $L_3\propto a^4$ is better perceivable at lower temperatures. In Fig. S2, we compare the $L_3$ for the $|\hspace{-2pt}\uparrow\rangle(-3,4)$ state at 80 nK with the 860 nK result presented in the main manuscript. 
  The Efimov resonance at about 140 G perturbs the overall scaling. At 150 G the $L_3$ value for 80 nK is again close to the $a^4$ prediction.

\end{document}